%% file: chartas.tex
\documentclass[apJ]{emulateapj}

\usepackage{textcomp}
\usepackage{epsfig, epstopdf}
\usepackage{amssymb,amsmath}

\newcommand{\simgt}{\lower 2pt \hbox{$\, \buildrel {\scriptstyle >}\over {\scriptstyle\sim}\,$}}
\newcommand{\simlt}{\lower 2pt \hbox{$\, \buildrel {\scriptstyle <}\over {\scriptstyle\sim}\,$}}

\slugcomment{Received 2013 May 16; Accepted 2013 July 21}
\shorttitle{An X-Ray Survey of Sub-DLAs}
\shortauthors{CHARTAS ET AL.}

\begin{document}

\def\sarc{$^{\prime\prime}\!\!.$}
\def\arcsec{$^{\prime\prime}$}
\def\beginrefer{\section*{References}%
\begin{quotation}\mbox{}\par}
\def\refer#1\par{{\setlength{\parindent}{-\leftmargin}\indent#1\par}}
\def\endrefer{\end{quotation}}

\title{A Mini X-Ray Survey of Sub-DLA\MakeLowercase{s}; Searching for AGN\MakeLowercase{s} Formed in Protogalaxies}

\author{G. Chartas\altaffilmark{1,2}, V. P. Kulkarni\altaffilmark{2},  and A. Asper\altaffilmark{1}}

\altaffiltext{1}{Department of Physics and Astronomy, College of Charleston, Charleston, SC, 29424, USA, chartasg@cofc.edu}

\altaffiltext{2}{Department of Physics and Astronomy, University of South Carolina, Columbia, SC, 29208, USA, kulkarni@sc.edu}

\begin{abstract}

\noindent

A significant fraction of the sub-damped Lyman-alpha (sub-DLA) absorption systems in quasar spectra appear to be metal-rich, 
many with even super-solar element abundances. This raises the question whether some sub-DLAs may harbor active galactic nuclei (AGN)
since supersolar metallicities are observed in AGN.  
Here we investigate this question based on a mini-survey of 21 quasars known to contain sub-DLAs in their spectra. 
The X-ray observations were performed with the {\sl Chandra X-ray Observatory}.
In cases of no detection we estimated upper limits of the X-ray luminosities of possible AGNs at the redshifts of the sub-DLAs.
In six cases we find possible X-ray emission within $\sim$ 1 \arcsec\ of the background quasar 
consistent with the presence of a nearby X-ray source.
If these nearby X-ray sources  are at the redshifts of the sub-DLAs, their estimated 0.2$-$10~keV luminosities range between
$0.8 \times 10^{44} \, h^{-2}$ and $4.2 \times 10^{44} \, h^{-2}$ erg s$^{-1}$, thus
%0.8$\times$10$^{44}$ h^{-2}$ and 4.2$\times$10$^{44} ,\$~h^{-2}$~erg~s$^{-1}$, thus 
ruling out a normal late-type galaxy origin, and suggesting that the emission originates in a galactic nucleus near the center of a protogalaxy.
The projected distances of these possible nearby X-ray sources from the background quasars
lie in the range of 3--7~$h^{-1}$~kpc, consistent with our hypothesis that they represent AGNs centered on the sub-DLAs. 
Deeper follow-up X-ray and optical observations are required to confirm the marginal detections of X-rays from these sub-DLA galaxies.

\end{abstract}


\section{INTRODUCTION}
The galaxy formation process is thought to be hierarchical with smaller dark matter haloes 
coalescing to form larger ones (e.g., for a review see Springel, Frenk \& White, 2006, and references therein).  
According to this theory  when the baryonic matter cools down enough to form molecular hydrogen
it falls into these haloes resulting in the birth of the first stars.
Eventually clusters of stars may form in the centers of these haloes. 
The merger of these haloes may produce the first protogalaxies.

The damped and sub-damped Lyman-alpha absorption line systems found in quasar and GRB spectra
are thought to be produced by intervening protogalaxies or young galaxies 
(e.g., Wolfe, Gawiser, \& Prochaska 2005, and references therein).
The damped Lyman-alpha (DLA) absorbers have neutral hydrogen  column densities log $N_{\rm HI}$ $ \geq $ 20.3, 
while sub-DLAs have 19.0 $ \leq $ log $N_{\rm HI}$ $ < $ 20.3. Together, DLAs and sub-DLAs constitute most of the 
neutral hydrogen in galaxies at high redshifts.  The nature of the protogalaxies associated with DLAs and sub-DLAs is very uncertain. 
%(e.g., Molaro \& Vladilo 1997).  
To understand where DLA/sub-DLAs fit in the big picture of galaxy evolution, it is of great interest to understand the nature of the galaxies associated with DLA/sub-DLA absorbers. 

Recent observations of element abundances show most DLAs to be metal-poor, while a substantial fraction 
of sub-DLAs appear to be metal-rich, i.e., near-solar or even super-solar; (see, e.g., Kulkarni et al. 2005, 
2007, 2010; Prochaska et al. 2003, 2006; Peroux et al. 2006; Meiring et al. 2008, 2009). 
A small fraction (about 5\%) of DLAs are also found to be metal-strong (Herbert-Fort et al. 2006). 
How did these metal-rich absorbers get so enriched $ > $ 7--10 Gyr ago? 
Supersolar metallicities are also found in AGN at both low and high redshifts, from emission-line observations (e.g., Storchi- Bergmann et al. 1998; Hamann et al. 2002; Dietrich et al. 2003; Nagao et al. 2006; Groves et al. 2006). The presence of AGN with supersolar metallicity at even $z \sim 5$ may seem surprising given that galaxies at high redshifts are expected to be more gas-rich, and less enriched compared to low-redshift galaxies (e.g., Kauffman \& Haehnelt 2000). The first major star formation episode in the high-redshift metal-rich AGN 
appears to have happened at $< 1$ Gyr after the Big Bang. In any case, the observations of high metallicities in AGN at high and low redshifts, together with the existence of metal-rich sub-DLA absorbers, raises the question of whether some of the metal-rich {\it absorbers} may be associated with galaxies possessing AGN  powered by super-massive black holes  (SMBH).

Indeed, the process by which supermassive black holes form and grow at the centers of galaxies is also very uncertain.
Proposed models of SMBH formation and growth (e.g., Kauffmann \& Haehnelt 2000; Wyithe \& Loeb 2003; Volonteri et al. 2003; Hopkins et al. 2006; Croton et al. 2006; Volonteri et al. 2013) include : (a) the direct collapse from dense cold gas clouds,
(b) contraction of dense cold gas into a supermassive
star that after fragmentation or multiple formation
events may form black hole binaries that spiral together via gravitational
radiation to form a SMBH,  (c) runaway growth of a massive  black hole by accretion, (d) mergers of black holes
and (e) more realistically, a combination of the above models.

It is unknown at what exact time during a galaxy's formation period the 
central massive black hole accretes enough gas to become active. 
A rough estimate of the duration of the active phase can be obtained from the observed fraction of 
galaxies that contain active nuclei. Specifically, surveys of galaxies indicate that about one in every hundred contain active nuclei implying AGN lifetimes of the order of  10$^{8}$ years.
It is thought that an AGN's lifetime will depend on the amount of time it takes for the accretion rate  to
drop below a critical level.  

Typical quasars and Seyfert galaxies require accretion rates of 
the order of a few  M$_{\odot}$ per year to  be active, whereas, the accretion rate onto a non-active nucleus such 
as SgrA* is $\sim$ 10$^{-9}$$-$10$^{-8}$ M$_{\odot}$ per year (e.g., Dexter et al. 2009; Shcherbakov et al. 2012).
There are various proposed processes that can lead to the reduction of the available gas reservoir that
feeds the central source. These include the ejection of gas from the galaxy by powerful AGN and starburst winds,
the formation of stars and the infall of gas into the central black hole. 

In protogalaxies these processes have not had enough time to deplete the gas reservoir 
and protogalaxies may contain enough gas to fuel a possible central massive black hole.
If this is the case, we might expect to find a large fraction of
protogalaxies with active nuclei. 
The protogalaxies sampled by metal-rich DLAs and sub-DLAs could thus be associated with SMBH with undepleted gas reservoirs. Examples of AGN in local analogues to protogalaxies
are the actively accreting massive black hole detected in the dwarf starburst galaxy Henize2-10
(Reines et al. 2011,2012) and AGN detected in local galaxies with properties
that are very similar to distant Lyman-break galaxies (Jia et al. 2011).

In \S 2 we present our sample of sub-DLAs and the analysis of X-ray observations; in
 \S 3 we discuss the possible detection of X-rays near the locations of background quasars that are known to contain sub-DLAs;
 and in \S4 we present a discussion of our results and prospects for expanding the sample and improving the confidence of the detections.
Throughout this paper we adopt a flat $\Lambda$ cosmology with 
$H_{0} = 70 \, h$ km s$^{-1}$ Mpc$^{-1}$, $\Omega_{\rm \Lambda}$ = 0.7, and  $\Omega_{\rm M}$ = 0.3.

\section{Sub-DLA Sample and X-ray Observations}

To test the plausibility that sub-DLAs may be associated with SMBHs
we performed an exploratory search for X-ray emission associated 
with a subset of the sub-DLAs presented in a survey of $ z > 4 $ quasars  (Guimaraes et al. 2009).
We also included five sub-DLAs detected in additional studies (Peroux et a., 2011; Prochaska et a. 2005; Som et al. 2013)

These sub-DLAs were chosen because X-ray spectra of their background quasars have been obtained with the
{\sl Chandra X-ray Observatory} (hereafter {\sl Chandra}) and are available 
in the {\sl Chandra X-ray Center}  archives.
Our search was restricted to {\sl Chandra} observations since only {\sl Chandra} has the 
spatial resolution of $\sim$ 0\sarc5; high spatial resolution is desirable to resolve X-ray emission
from the candidate protogalaxies associated with the sub-DLAs.
As a result of this search 
we found 21 quasars with sub-DLAs that had been
observed with the Advanced CCD Imaging Spectrometer (ACIS; Garmire 2003) on board 
{\sl Chandra}. 
A log of the X-ray observations that includes observation dates, observational identification numbers, exposure times, ACIS frame time and the observed 0.2$-$10~keV counts is presented in Table~1. \\

\section{X-ray Mini-Survey of Sub-DLAs}
The {\sl Chandra} observations of the sub-DLAs of our mini-sample were analyzed using the standard software CIAO 4.5 
provided by the {\sl Chandra X-ray Center} (CXC). 
We used standard CXC threads to screen the data for status, grade, and time intervals of acceptable aspect solution and background levels. 
To improve the spatial resolution we employed the sub-pixel resolution technique developed by Li et al. (2004) and incorporated via the
Energy-Dependent Subpixel Event Repositioning (EDSER) algorithm into the tool acis\_process\_events of CIAO 4.5.
For comparison we also employed the sub-pixel resolution technique developed by Tsunemi et al. (2001).
In 17 out of 21 cases the background quasar is detected with {\sl Chandra}.
In 6 out of the 21 {\sl Chandra} observations 
we detected X-ray emission (resolved from the quasar itself) within $\sim$1~\arcsec\ of the background
quasar that is known to contain a sub-DLA  in its optical/UV spectrum. 
One exciting possibility is that 
this X-ray emission originates from an AGN near the center of a protogalaxy.
In Figure 1 we show the {\sl Chandra} images of the fields 
around quasars PSS~0121$+$0347, PSS~0133$+$0400, PSS~0955+5940, SDSS~J095744.46$+$330820.7, Q~1323$-$0021, and PSS~2322+1944, near which we detected X- ray emission possibly associated with the intervening systems.
The {\sl Chandra-ACIS} Point Spread Function (PSF) has a FWHM of $\sim$ 0\sarc5, and is indicated with the solid circles centered 
on the optical positions of the quasars. We detect X-rays centered on the background quasars, however, we also detect a significant number of counts asymmetrically distributed and located about 0.5 $-$1\arcsec\ from the optical positions of the quasars. 
Regions containing $\simgt$ 3  events with a distance of greater than 0.5 arcsec from the background quasar
and clustered within a 0.5 arcsec radius circle were considered as marginally detected
X-ray sources.
The dashed circles are centered on the marginally detected  X-ray emission within 
$\sim$ 1 \arcsec\ of the background quasars.
In Table 2 we list the properties of the background quasars and their sub-DLAs.

The spatial distributions of the excess X-ray counts near the quasars are consistent with being produced by
point sources in the sense that these counts are clustered within circles of radii of less than 0\sarc5 (ACIS PSF FWHM $\sim$ 0\sarc5). X-ray emission from the host galaxy is too weak to be detected at the redshift of the quasar. 
PSS 2322+1944 is a gravitationally lensed quasar with the second lensed image represented with a circle near label B in Figure 1. 
We find a marginal detection of X-ray counts near the location of the second lensed image.
We note that the background in these {\sl Chandra} observations is very low as evident with the very
few counts detected away from these sources.  
In Figure 2 we also show the observed surface brightness profiles centered on the quasars Q~1323-0021 and PSS 2322$+$1944
and compare them to simulated ones for a point source. The other four candidates have too few counts to produce
surface brightness profiles.
We employed the MARX tool to simulate PSFs. The observed surface brightness profiles deviate from those simulated 
for a single point source and the presence of additional sources within 1\arcsec\ of the quasar is
consistent with our image analysis. We note that these quasars were centered on-axis at the aim-point of ACIS S3.

In the case of the moderate S/N image of Q~1323-0021, the detected counts were sufficient for a two-dimensional 
fit to both the background quasar and nearby X-ray sources with simulated PSFs.
The image was binned with a bin size of 0\sarc0246 bins (compared to the 0\sarc491 ACIS pixel scale) 
and fit by minimizing the $C$-statistic (Cash 1979) between the observed and model images.
In Figure 3 we show the best-fit PSF model of the {\sl Chandra} observation of Q~1323-0021. 
The positions of the PSFs were left as free parameters in the fit and the best-fit values of the 
positions of the nearby sources labelled as S1 and S2 are listed in Table 3. 
S1 and S2 lie at angular separations of 0\sarc47 and 0\sarc43\, respectively from the background quasar. Neither S1 nor S2 is seen in the K$^{\prime}$-band adaptive optics image with FWHM 0\sarc08 obtained by Chun et al. (2010). However, we note that a massive candidate galaxy was reported by Chun et al. (2010) to be located 
1\sarc25 away from the quasar. It is possible that S1 and S2 are not bright enough in near-IR, or they are lost in the artifacts of the AO PSF in the image of Chun et al. (2010). 
We note that the intervening absorber of Q~1323-0021 is one of the highest metallicity absorbers known. Specifically, 
P{\'e}roux et al. (2006) reported metallicities of [Zn/H] = +0.61 $\pm$ 0.20 and [Fe/H] = $-$0.51 $\pm$ 0.20.

\section{Discussion and Conclusions}

Our exploratory mini-survey of sub-DLAs  contained mostly short snapshot {\sl Chandra} observations
with exposure times ranging between 2-19~ks. In 17 of the 21 {\sl Chandra} observations of 
of the sub-DLA fields in our sample
we detect X-rays from 
the background quasars that are known to contain the sub-DLAs.
In  the  {\sl Chandra}  observations of six quasars of our sample
we also have marginal detections of X-ray sources within 1\arcsec\ of the background quasars.
These X-ray sources may be associated with the sub-DLAs.
We briefly discuss possible origins of these nearby X-ray sources.

(a) X-ray emission from a nearby X-ray source instead of the sub-DLA.\\
We estimated the probability of detecting a background X-ray source within a circle of radius 1~\arcsec\ centered on the quasar
by scaling the cumulative number counts per square degree found in the 4 Ms Chandra Deep Field-South survey (Lehmer et al. 2012) by the ratio of the areas. 
In Table 3 we list the 0.5--8~keV fluxes of the candidate sub-DLA sources and the probability of finding by chance a second source with 
a flux greater than or equal to the one detected. We conclude that it is unlikely that our six candidate X-ray sources are 
random background X-ray sources in the sky.
We are being conservative in these estimates since the probability of detecting a faint source near a bright quasar is even smaller than our 
estimated values.

(b) X-ray emission from normal late-type galaxies instead of the sub-DLA.\\
X-rays from normal galaxies are thought to be produced by supernovae, supernova remnants, stellar outflows of hot gas, the 
hot interstellar medium, high and low-mass X-ray binaries and young stars (e.g., Fabbiano 2006).
An analysis of normal late-type galaxies detected in the {\sl  Chandra Deep Field} survey by Lehmer et al (2012) indicates that their
0.5$-$8~keV luminosities increase on average from 
$\sim 3 \times 10^{39}~h^{-2}$ erg s$^{-1}$ at $z = 0.1$  to 
$\sim 2 \times 10^{42}~h^{-2}$ erg s$^{-1}$ at $z = 1.4$
and their mean X-ray luminosity to star formation rate is found to be constant within this redshift span. 
An analysis of normal galaxies in the Great Origins Deep Survey fields by Ptak et al. (2007) shows that their X-ray evolution 
can be expressed as L$^{*}$($z$) = (1+$z$)$^{p}$L$^{*}$(z=0), where $p = 1.6$ for early-type and $p = 2.3$ for late-type galaxies consistent with what is 
found in the independent survey of Lehmer et al. (2012). 
Assuming the nearby X-ray sources detected in our mini-survey are located at the redshifts of the sub-DLAs,
their estimated X-ray luminosities (see Figure 3) are significantly larger than what would be estimated if they were originating from normal late-type galaxies at those redshifts.

(c) X-ray emission from AGN associated with the sub-DLAs. \\
One exciting possibility is that the X-ray emission of the nearby sources of our mini-survey originates from AGN located 
near the center of the protogalaxy. 
Assuming the X-ray emission of the nearby sources
originates at the redshifts of the intervening sub-DLAs 
we estimate their 0.2$-$10 keV luminosities to range between  
$0.8 \times 10^{44} \, h^{-2}$ and $4.2 \times 10^{44} \, h^{-2}$ erg s$^{-1}$
consistent with that of a low-luminosity AGN. 
For estimating the luminosities we assumed a power-law model with a photon index of 
$\Gamma = 1.8$ modified by Galactic absorption. 
The angular separation of these X-ray sources from the  background quasars correspond
to projected linear separations at the redshift
of the sub-DLA in the range of 3--7~$h^{-1}$~kpc.  These separations are consistent with the expected impact-parameters
of the background quasars from the center of the sub-DLAs at the redshift of the foreground absorbers.
In Table 3 we provide several properties of the 
candidate X-ray sources, assuming they are at the redshift of the sub-DLAs.

(d) X-ray emission from an additional image produced by gravitational lensing of the background quasar. \\
The probability of a quasar being gravitationally lensed into multiple images
depends primarily on the comoving number density of lenses and the lensing cross section
of each lens and is of the order of 0.1\% (e.g, Turner et al. 1984; Comerford et et al. 2002).
The fraction of background quasars in our sample with a detected candidate nearby X-ray source (6/21) is
significantly larger than the expected number of gravitationally lensed quasars in our sample.
The probability therefore that any of the quasars in our sample is gravitationally lensed (in addition
to PSS 2322 that is a known gravitationally lensed quasar) is negligible.
We note that gravitationally lensed quasars are often included in surveys of
intervening absorption systems because of
their advantage of having multiple lines of sight through intervening absorbers.
We searched catalogs of known gravitationally lensed systems
(e.g., Master Lens database of the Orphan Lenses Project) and found that none of our
candidate X-ray sources are listed as lensed quasars other than PSS 2322.

(e) X-ray emission from intense star formation associated with the sub-DLAs or the host galaxies of the background quasars. \\
Lehmer et al. (2010) determine the relation between star formation rate (SFR) and 2--10~keV luminosity (L$^{Gal}_{\rm X}$)
for a sample of galaxies composed of normal galaxies, luminous infrared galaxies (LIRGS) and 
ultraluminous infrared galaxies ULIRGS to be 
 log(L$^{Gal}_{\rm X}$) = $\alpha$ + $\beta$log(SFR),
where  $\alpha$ = 39.49 $\pm$ 0.21,  $\beta$ = 0.74 $\pm$ 0.12, and SFR is in units 
of $\rm M_{\odot}~yr^{-1} $.
We find that the observed 2--10~keV luminosities (listed in Table 3) of the possible X-ray sources near the background quasars 
are several orders of magnitude larger than the inferred L$^{Gal}_{\rm X}$ values 
assuming the nearby X-ray emission originates from intense star formation associated with the sub-DLAs or the host galaxies of the background quasars.

Follow-up observations with longer exposures are required to confirm or refute the discovery of a
possible new class of X-ray sources, in particular ones that may be associated with active protogalactic nuclei.
Specifically, spectral and variability analyses of deeper X-ray observations of these sources
will increase the S/N of these detections and provide insight into their nature.
Follow-up  deep spectroscopic and imaging optical and UV observations
may  also provide the redshifts and images of possible galaxies at the locations of the X-ray sources.

\acknowledgments
We acknowledge financial support from NASA via the Smithsonian Institution grant SAO AR0-11019X.
VPK acknowledges partial support from 
the National Science Foundation grant AST/1108830.

\clearpage
\input{table1.tex}

\clearpage
\input{table2.tex}

\clearpage
\input{table3.tex}

  \begin{figure}
   \includegraphics[width=15cm]{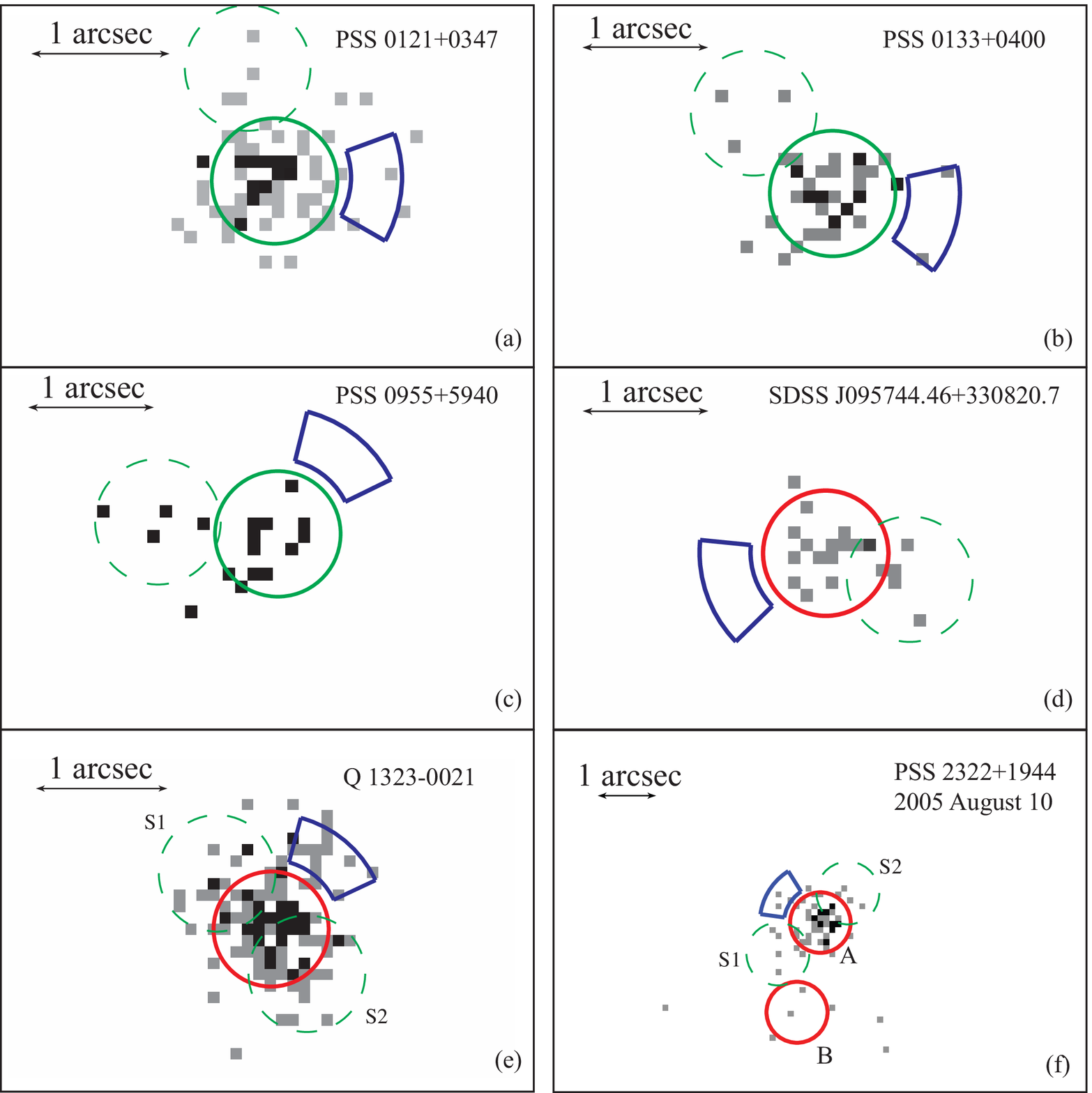}
        \centering
\caption[]{
{\sl Chandra} images of the fields near quasars PSS~0121$+$0347 (a), PSS~0133$+$0400 (b), PSS~0955+5940 (c), SDSS~J095744.46$+$330820.7 (d), Q~1323$-$0021 (e), and PSS~2322+1944 (f). 
The solid circles are centered on the background quasars
and have radii of 0\sarc5, which is approximately the size of the point spread function of {\sl Chandra-ACIS} on axis.
The images suggest the presence of X-ray sources within $\sim$ 1 \arcsec\ of the quasars.
 The dashed circles are centered on possible nearby X-ray sources.
PSS~2322$+$1944 is a known gravitationally lensed quasar with images A and B
separated by $\sim$ 1\sarc5.  Background noise in these
{\sl Chandra} observations is negligible as evident from the lack of events just a few arcsec away
from the central sources. 
The  X-ray sources we report (denoted as dashed circles) are not near the locations of the PSF artifacts (denoted as pie-shaped regions).
North is up and east is to the left. }
\label{fig:stacked}
\end{figure}

\clearpage

  \begin{figure}
   \includegraphics[width=15cm]{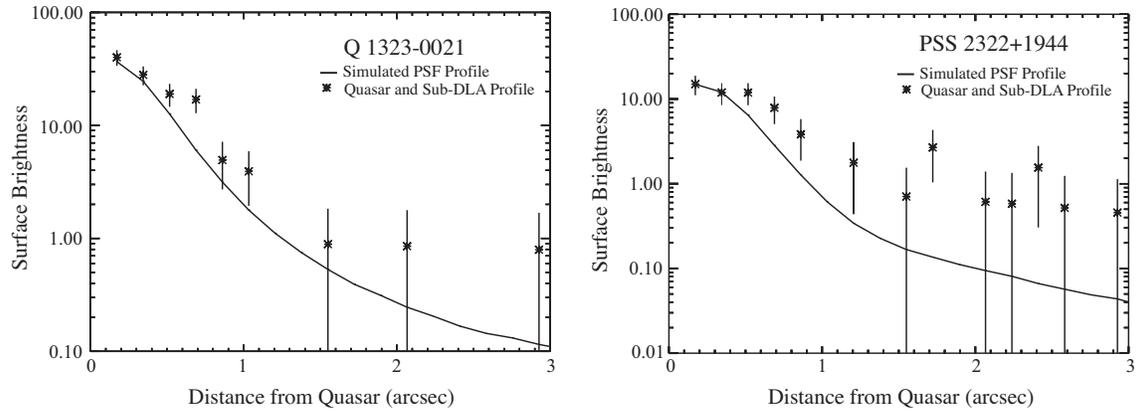}
        \centering
\caption[]{
Observed surface brightness profiles of the X-ray emission
centered on quasars  Q~1323$-$0021 and  PSS~2322$+$1944. 
For PSS~2322$+$1944 we stacked the images from the 2002 and 2005 observations.
The solid lines represent simulated surface brightness profiles of point sources. 
The observed X-ray surface brightness profiles deviate significantly from those of point
sources at distances of about 0\sarc5 from the quasar, 
consistent with the presence additional nearby point sources as
shown in Figure 1. In the case of PSS~2322+1944 the observed deviations at distances larger than  0\sarc5
are possibly due to lensed image B located about 1\sarc5  from the brighter image A.}  
\label{fig:stacked}
\end{figure}

\clearpage

  \begin{figure}
   \includegraphics[width=15cm]{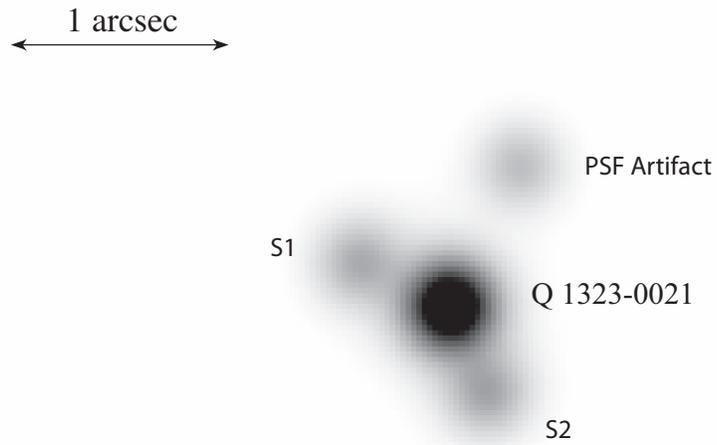}
        \centering
\caption[]{ Best-fit PSF model to the {\sl Chandra} observation of Q~1323$-$0021. 
North is up and east is to the left.
}
\label{fig:stacked}
\end{figure}

\clearpage

  \begin{figure}
   \includegraphics[width=15cm]{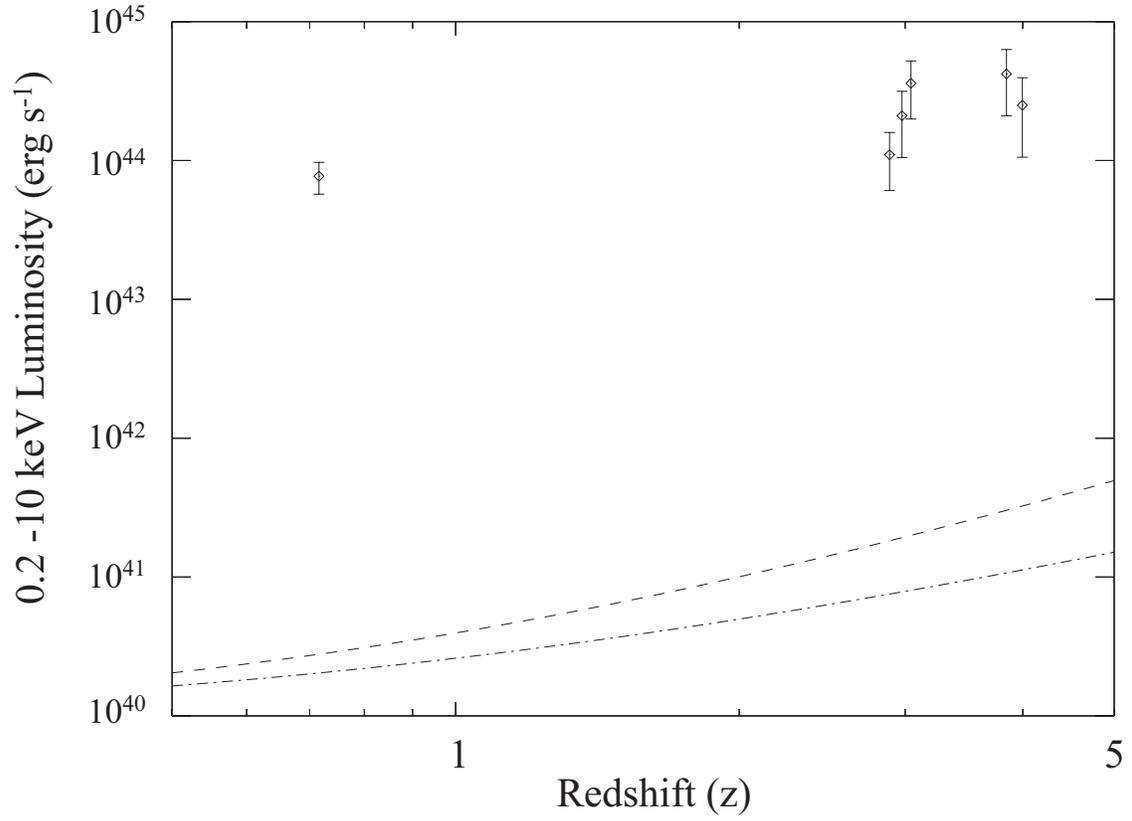}
        \centering
\caption[]{
0.2$-$10~keV luminosities of candidate X-ray sources possibly associated with sub-DLAs versus redshift. The dashed and dot-dashed lines represent the predicted X-ray luminosity 
evolution of normal late-type and early-type galaxies, respectively,  presented in Ptak et al. (2007) and consistent with the survey of Lehmer et al. (2012). }
\label{fig:subdlas}
\end{figure}

\end{document}

%% file: table1.tex
\begin{deluxetable}
{llccclll} \tabletypesize{\scriptsize}

\tablecolumns{8} \tablewidth{0pt} \tablecaption{
Log of Observations  \label{tab:obid}}

\tablehead{
 \multicolumn{1}{c}{} & \multicolumn{1}{c}{Chandra} & \multicolumn{1}{c}{Chandra} &
\multicolumn{1}{c}{} & \multicolumn{1}{c}{} & \multicolumn{1}{c}{} & \multicolumn{1}{c}{} &\multicolumn{1}{c}{}  \\
\multicolumn{1}{c}{Object Name} & \multicolumn{1}{c}{Observation} & \multicolumn{1}{c}{Observation} &
\multicolumn{1}{c}{RA(2000)$^a$} & \multicolumn{1}{c}{Dec(2000)$^a$} & \multicolumn{1}{c}{$t_{\rm exp}$$^b$} & \multicolumn{1}{c}{N$^c$} &\multicolumn{1}{c}{Ref.}  \\
 \multicolumn{1}{c}{} & \multicolumn{1}{c}{Date} & \multicolumn{1}{c}{ID} &
\multicolumn{1}{c}{} & \multicolumn{1}{c}{} & \multicolumn{1}{c}{(s)} & \multicolumn{1}{c}{(counts)} &\multicolumn{1}{c}{}  \\
}

\startdata
PSS 0121+0347                                     & 2002 February 07     & 3151    & 01 21 26.1 & $+$03 47 07.0 &  5684 &  76 $\pm$ 9 & 1 \\
PSS 0133+0400                                     & 2001 November  26    & 3152    & 01 33 40.3 & $+$04 00 59.0  & 6076 & 42 $\pm$ 7 & 1 \\ 
PSS 0209+0517                                     & 2002 January 14          & 3153   & 02 09 44.7 & $+$05 17 14.0  & 5776 & 28 $\pm$ 6  & 1\\ 
SDSS J074749.74+443417.1              & 2002 December 24      & 4068   & 07 47 49.7 & $+$44 34 16.0  & 4534 & 9  $\pm$  3 & 1\\  

SDSS075618+410408                          &  2002 February 08       & 3032   & 07 56 18.1 & $+$41 04 08.6     &  7327     & 21  $\pm$  5 & 1\\ 

PSS 0955+5940                                      & 2002 April 14                 & 3156   & 09 55 11.3 & $+$59 40 31.0  & 5681 & 18 $\pm$ 4 & 1\\  % structure!
SDSS J095744.46+330820.7               & 2001 December 24       & 3157   & 09 57 44.4 & $+$33 08 21.0  & 6035 & 20 $\pm$ 5  & 1\\ 
PSS 1057+4555                                       & 2000 June 14                & 878      & 10 57 56.3 & $+$45 55 53.0  & 2807 & 37 $\pm$ 6 & 1\\   % structure ?

SDSS J122556.61+003535.1                  &  2006 June 25               & 6877    & 12 25 56.6 & $+$00 35 35.1  &  2885 & 9 $\pm$ 3  & 2 \\
SDSS J124942.12+334953.8                 & 2001 March 24               & 2084   & 12 49 42.2 &$+$33 49 54.0  & 4619 & 20 $\pm$ 5 & 1\\ 

SDSS J131743.12+353131.8                  & 2000 June 14                 & 879      & 13 17 43.1 & $+$35 31 31.8  & 2787 & 7 $\pm$ 3  & 3\\ 

Q1323-0021                                                 & 2004 May 09                &  4855     &  13 23 23.7  & $-$00 21 55.2  & 4360 & 151 $\pm$ 12 &  2\\ 

SDSS J132512.49+112329.7                 & 2004 March 12              & 3565     & 13 25 12.5 & $+$11 23 29.7  & 4701 & 32 $\pm$ 6 & 1 \\
SDSS J132611.85+074358.4                & 2002 January 10           & 3158     & 13 26 11.9 &$+$07 43 58.0  & 5886 & 72 $\pm$ 9 &  1\\ 
                                                                      & 2011 March 07             & 12794     &     &   & 4982 & 43 $\pm$ 7 &  1\\

PSS 1435+3057                                         & 2000 May 21                 & 880      &  14 35 23.0  & $+$30 57 22.0  & 2881 & $<$ 5 & 1\\

PSS 1646+5514                                        & 2003 September 09      & 4072     & 16 46 56.3 & $+$55 14 45.0  & 4843 & 7 $\pm$ 3  & 1 \\
SDSS 173744+582829                           & 2002 August 05              & 3038     & 17 37 44.9 & $+$58 28 30.0  & 4616 & $<$ 5  & 1 \\  %
SDSS J2123-0050                                    &  2006 March 30              & 6822      & 21 23 29.5  & $-$00 50 53.0   & 3904  &  26 $\pm$ 5  & 4\\ % structure ?       
PSS 2322+1944                                      & 2002 August 23              & 3028      & 23 22 07.1 & $+$19 44 23.0  & 4897 & 17 $\pm$ 4 &  1\\ % structure
                                                                   & 2005 August 10              & 5605      &   &  & 13600 & 66  $\pm$ 8 & 1\\ % structure
PSS J2344+0342                                     & 2003 November 20       & 4074       & 23 44 03.2 & $+$03 42 26.0  & 5099 & $<$ 5 &  1\\

SDSS J235253.51-002850.4                & 2002 May 16                   & 2115      & 23 52 53.5   & $-$00 28 50.4 & 5779& $<$ 5 &  2 \\

\enddata

\tablenotetext{a}{Optical positions in J2000.0 equatorial coordinates.}

\tablenotetext{b}{Effective exposure time after data processing.}

\tablenotetext{c}{Background subtracted source counts for events with energies in the 0.2$-$10 keV band in a circular region of radius  2\sarc5 centered on the quasar.}

\tablerefs{ (1) Guimaraes et al. 2009; (2)  Peroux et al. 2011; (3) Prochaska et al. 2005;  (4) Som et al. 2013} %(5) Rao et al. 2006: Jackson et al. 2008.}

\end{deluxetable}

%% file: table2.tex
\begin{deluxetable*}
{llccccc} \tabletypesize{\scriptsize}

\tablecolumns{7} \tablewidth{0pt} \tablecaption{
Properties of Background Quasars and Sub-DLAs \label{tab:obid}}

\tablehead{
\multicolumn{1}{c}{Object Name} & \multicolumn{1}{c}{$z_{\rm em}$} & \multicolumn{1}{c}{$z_{\rm abs}$} &
\multicolumn{1}{c}{$\log$N$_{\rm G,H}$$^a$} & \multicolumn{1}{c}{$\log$N$_{\rm abs,H~I}$$^b$} & \multicolumn{1}{c}{L$_{\rm qso}$$^c$} & \multicolumn{1}{c}{L$_{\rm subDLA}$$^d$}  \\
\multicolumn{1}{c}{} & \multicolumn{1}{c}{} & \multicolumn{1}{c}{} &
\multicolumn{1}{c}{cm$^{-2}$} & \multicolumn{1}{c}{cm$^{-2}$} & \multicolumn{1}{c}{($10^{45}~h^{-2}$~erg~s$^{-1}$)} & \multicolumn{1}{c}{($10^{44}~h^{-2}$~erg~s$^{-1}$)}  \\
}

\startdata

PSS 0121+0347                                     & 4.130    & 2.977  & 20.544 & 19.50 $\pm$ 0.15 &  7.1(8.3)  & 2.1 \\
PSS 0133+0400$^{e}$                          & 4.154    & 3.995  & 20.491 & 20.15 $\pm$ 0.10 & 3.2(4.2) &  2.5   \\ 
PSS 0209+0517$^{e}$                          & 4.174    & 3.862  & 20.668 & 20.30 $\pm$ 0.10 & 2.2(2.9)  & $<$ 4.0 \\ 
SDSS J074749.74+443417.1               & 4.432    & 3.139  & 20.719 & 20.00 $\pm$ 0.20 &  0.84(1.7)  &$<$ 4.0\\  
SDSS075618+410408                           & 5.09     & 4.360  & 20.685 & 20.15 $\pm$ 0.10 & 1.9(1.8)  &$<$ 3.8   \\ 
PSS 0955+5940                                     & 4.336    & 3.843  & 20.134 & 20.00 $\pm$ 0.15 & 2.1(1.7)  &  4.2 \\  % structure!
                                                                   &              & 4.044  &                 & 20.10 $\pm$ 0.15 &                                           & 4.7 \\  % structure!
SDSS J095744.46+330820.7               & 4.227   & 3.043  & 20.179 & 19.65 $\pm$ 0.15 &  2.6(2.4) & 3.6   \\ 
                                                                   &                & 3.364  &               & 19.70 $\pm$ 0.15        & & 4.4    \\ 
                                                                    &          & 3.900  &        & 19.50 $\pm$ 0.10        & &  5.8    \\ 
PSS 1057+4555                                     & 4.137    & 2.909  & 20.061 & 20.05 $\pm$ 0.10  &  5.5(6.4) & $<$ 5.1   \\   % structure ?
                                                                   &          & 3.058  &        & 19.80 $\pm$ 0.15  &  &  $<$ 5.6\\   % structure ?
                                                                  &          & 3.164  &        & 19.50 $\pm$ 0.20 &  &  $<$ 6.0 \\   % structure ?
                                                                  &          & 3.317  &        & 20.15 $\pm$ 0.10 &  & $<$ 6.6 \\   % structure ?
SDSS J122556.61+003535.1             & 1.226    & 0.773 & 20.283 & 21.38 $\pm$ 0.12 & 0.18(0.37)   &$<$ 0.36 \\
SDSS J124942.12+334953.8              & 4.897    & 4.572  & 20.097 & 19.80 $\pm$ 0.10 & 4.3(4.2) &$<$ 11  \\ 
SDSS J131743.12+353131.8             & 4.381    & 3.461  & 19.998 & 19.90 $\pm$ 0.05  &1.1(2.1)   &$<$ 9.3   \\ 
Q1323-0021                                           & 1.388    & 0.716  & 20.270 & 20.21 $\pm$ 0.20 & 2.2(2.2)   & 0.5(S1),0.5(S2)    \\ 
SDSS J132512.49+112329.7           & 4.400    & 3.723  & 20.281 & 19.50 $\pm$ 0.20 & 4.9(4.1) &$<$ 5.7  \\
                                                                &          & 4.133  &        & 19.50 $\pm$ 0.20 &  &$<$ 6.9  \\
SDSS J132611.85+074358.4$^{e}$           & 4.123    & 2.919  & 20.303 & 19.95 $\pm$ 0.10  & 6.7(6.9)  &$<$ 2.5   \\ 
                                                               &          & 3.425  &        & 19.90 $\pm$ 0.15  &  & $<$ 3.5  \\ 

SDSS J132611.85+074358.4$^{f}$           & 4.123    & 2.919  &20.303    & 19.95 $\pm$ 0.10  & 6.5(5.8) &$<$ 3.9   \\ 
                                                               &          & 3.425  &        & 19.90 $\pm$ 0.15  &  & $<$ 5.3   \\

PSS 1435+3057                                & 4.350    & 3.267  & 20.068 & 20.05 $\pm$ 0.10 & $<$1.3  & $<$ 7.5    \\
                                                              &          & 3.516  &        & 20.20 $\pm$ 0.10 &   &$<$   8.7 \\
                                                              &          & 3.778  &        & 19.85 $\pm$ 0.10 &    & $<$  10.0 \\
PSS 1646+5514                                & 4.084    & 2.932  & 20.382 & 19.50 $\pm$ 0.10 & 0.95(0.78) & $<$ 5.1   \\
                                                             & 4.084    & 4.029  &        & 19.80 $\pm$ 0.15 &  & $<$ 9.3    \\
SDSS 173744+582829                   & 4.940    & 4.152  & 20.550 & 19.85 $\pm$ 0.15  & $<$ 8.1   &  $<$ 66.0  \\  %

SDSS J2123-0050                          & 2.261    & 2.058  & 20.667 & 19.35 $\pm$ 0.10 & 1.2(1.4)& $<$ 2.2    \\ % structure ?       
PSS 2322+1944$^{g}$                              & 4.118    & 2.888  & 20.655 & 19.95 $\pm$ 0.10 & 2.0$\mu^{-1}_{X}$(1.8$\mu^{-1}_{X}$) & $<$ 3.0   \\ % structure
                                                            &          & 2.975  &        & 19.80 $\pm$ 0.10  &  &  $<$ 3.2\\ % structure
PSS 2322+1944$^{h}$                               & 4.118    & 2.888  & 20.655 & 19.95 $\pm$ 0.10 & 2.7$\mu^{-1}_{X}$(2.6$\mu^{-1}_{X}$) &  1.1(S1),0.89(S2)   \\ % structure
                                                            &          & 2.975  &        & 19.80 $\pm$ 0.10  &  &  1.2(S1),0.96(S2) \\ % structure
PSS J2344+0342                            & 4.239    & 3.884  & 20.735 & 19.80 $\pm$ 0.10 & $<$ 0.9 & $<$ 0.8  \\

SDSS J235253.51-002850.4        & 1.628    & 0.873  & 20.543 & 19.18 $\pm$ 0.10 & $<$ 0.09 & $<$0.2  \\
                                                             &          & 1.032  &               & 19.81 $\pm$ 0.13 &  & $<$ 0.3\\
                                                             &          & 1.247  &               & 19.60 $\pm$ 0.25 &  &  $<$ 0.5\\

%LBQS 2350-0045B$^{h}$                        & 0.764     & 0.6044 & 20.539 &  21.54 0.15  &  &  \\  Gravitational Lens

\enddata

\tablenotetext{a}{Total Galactic hydrogen column density from Dickey \& Lockman (1990) }

\tablenotetext{b}{H~I column density.}

\tablenotetext{c}{0.2$-$10~keV luminosity of background quasar assuming a spectral model consisting of a power-law with a photon index fixed at $\Gamma=1.8$ and Galactic absorption. The values listed in parenthesis represent  0.2$-$10~keV luminosities obtained from spectral fits assuming the same model where the photon index was allowed to vary.}  

\tablenotetext{d}{Upper limit of the 0.2$-$10~keV luminosity of the sub-DLA assuming a 5~count detection within a circle of radius 0\sarc5. In cases with detected nearby sources we list the observed 0.2$-$10~keV luminosities of the subDLA's. In cases where two sources are detected
near the background quasar, the luminosity of each source is followed by its label shown in Figure 1.}

\tablenotetext{e}{The estimated luminosities correspond to the 2002 January 10 observation of SDSS J132611.85+074358.4}

\tablenotetext{f}{The  estimated luminosities correspond to the 2011 March 7 observation of SDSS J132611.85+074358.4}

\tablenotetext{g}{The estimated luminosities correspond to the 2002 August 20 observation of PSS~2322+1944. 
We have assumed that  the model-predicted magnification in the optical band of $\mu_{opt}$ $\approx$ 4.7,  
(Riechers et al. 2008) is equal to the X-ray magnification, $\mu_{X}$.}

\tablenotetext{h}{The estimated luminosities correspond to the 2005 August 10 observation of PSS~2322+1944.}

\end{deluxetable*}

%% file: table3.tex
\begin{deluxetable*}
{lccccccccc} \tabletypesize{\scriptsize}

\tablecolumns{10} \tablewidth{0pt} \tablecaption{
Properties of Candidate X-ray Sources \label{tab:obid}}

\tablehead{
\multicolumn{1}{c} {Quasar} &
\multicolumn{1}{c} {$z_{\rm em}$} &
\multicolumn{1}{c} {$z_{\rm abs}$} &
\multicolumn{1}{c} {log$N_{\rm H I}$${}^{a}$}  &
\multicolumn{1}{c} {$\Delta$x${}^{b}$} &
\multicolumn{1}{c} {$\Delta$y${}^{b}$} &
\multicolumn{1}{c} {D${}^{c}$} &
\multicolumn{1}{c} {$f_{\rm 0.2-10}$${}^{d}$} &
\multicolumn{1}{c} {$L_{\rm 2-10}$${}^{e}$} &
\multicolumn{1}{c} {$P$${}^{f}$} \\
\multicolumn{1}{c} {} &
\multicolumn{1}{c} {} &
\multicolumn{1}{c} {} &
\multicolumn{1}{c} {(cm$^{-2}$)} &
\multicolumn{1}{c} {(\arcsec)} &
\multicolumn{1}{c} {(\arcsec)} &
\multicolumn{1}{c} {(kpc)} &
\multicolumn{1}{c} {} &
\multicolumn{1}{c} {} &
\multicolumn{1}{c} {10$^{-4}$} \\
}

\startdata
%PSS 2322+1944                 & 4.17  & 2.98  & 19.8& -0.72 $\pm$ 0.02   & 0.0 $\pm$ 0.1      &  4.9       & 0.4     & 2.0$\times$10$^{44}$      \\
%LBQS 2350-0045B            & 0.76  & 0.60  & 21.5 &-0.77 $\pm$ 0.03    & 0.1 $\pm$ 0.1       & 5.2       & 4.9    & 5.3$\times$10$^{43}$      \\
%LBQS 2350-0045B            &  0.76 &  0.60 & 21.5  & 0.0 $\pm$ 0.1       & -0.8 $\pm$ 0.1       & 5.9       &  2.4    &       2.7$\times$10$^{43}$              \\
%PSS 0955+5940 & 4.34    & 3.84                & 20.0 & -2.0 $\pm$ 0.2     & -0.2 $\pm$ 0.2     & 13.6       & 0.7      & 3.5$\times$10$^{44}$     \\
%PSS 1057+4555 & 4.13   & 3.06                 &19.8  &  0.9 $\pm$ 0.2     & -0.2 $\pm$ 0.2     & 6.3         &  3.6     & 1.0$\times$10$^{45}$    \\
PSS 0121+0347                       & 4.130  & 2.977  & 19.50  & $+$0.28    & $+$0.68      &  5.71       &   3.84    &  1.8  & 2.4 \\
PSS 0133+0400                       &4.154   & 3.995  & 20.15  &  $+$0.72    &  $+$0.73     &  7.16       &   2.94    &  2.2   & 2.9  \\
PSS 0955+5940                       &4.336   & 3.843  & 20.00  &  $+$0.58     & $+$0.19      &  4.31       &   4.80    & 3.4    &  1.9\\
%                                                      &              & 4.044  & 20.00  &             &               & 4.36        &               & 4.70     & \\
PSS 2322+1944 (S1)                        &4.118    & 2.888   &19.95   & $+$0.72    & $-$0.50      & 6.85        &  2.58     & 0.94 & 3.1  \\
 %                                                     &               & 2.975  & 19.80  &              &               & 7.00       &               &     0.94  &\\
PSS 2322+1944 (S2)                       &               &            &    & $-$0.50     & $+$0.38      & 4.90        & 2.09              & 0.75  & 3.9  \\
 %                                                     &              &2.975   & 19.80  &               &               & 5.02        &              &     0.94   & \\
Q 1323-0021 (S1)                             & 1.388   & 0.716   &20.21   &$+$0.42      & $+$0.22      & 3.39        &  33.3    & 0.3  &   0.17\\
Q 1323-0021 (S2)                             &              &              &      &$-$0.17      & $+$0.39      & 3.10        & 33.3       &  0.3          &   0.17\\
 %                                                     &              &              &              &$-$0.25      &$-$0.37       & 3.29        &           & 53.5      & \\
SDSS J095744.46+330820.7 & 4.227  & 3.043   &19.65   & $-$0.44      & $-$0.34      & 4.29        &   6.4    & 2.7     & 1.4 \\
%                                                      &              & 3.364  & 19.70  &                &                & 4.29        &              & 4.40    & \\
 %                                                     &              & 3.900  & 19.50  &                &                &  4.06        &              & 5.80     & \\

\enddata

\tablenotetext{a}{Hydrogen column density of sub-DLA.}

\tablenotetext{b}{The RA and Dec offsets of the centroids of the candidate X-ray sources detected near the quasars. In cases where two nearby sources 
are detected they are labelled as S1 and S2 (see Figure 1).}

\tablenotetext{c}{The projected distances of these candidate X-ray sources from the quasars (scale lengths calculated at $z_{\rm abs}$).}

\tablenotetext{d}{The observed 0.2--10~keV fluxes of these candidate X-ray sources in units of 10$^{-15}$erg~s$^{-1}$~cm$^{-2}$ .}

\tablenotetext{e}{Absorbed 2--10~keV luminosity for a source at  $z_{\rm abs}$
in units of 10$^{44}$~$h^{-2}$~erg~s$^{-1}$. To estimate the luminosity we have assumed a power-law model with a photon index 
fixed at $\Gamma = 1.8$ modified by Galactic absorption.}

\tablenotetext{f}{The probability of finding by chance a
 background X-ray source within a circle of radius 1 \arcsec\ centered on the quasar
with a flux greater than or equal to the one detected.}

\end{deluxetable*}